\documentclass[onecollarge]{svjour2}
\usepackage{epsfig}
\usepackage{amsmath}
\usepackage{amssymb}
\usepackage{graphicx}
\usepackage{color}
\usepackage{bm}

\journalname{Few-Body Syst}

\begin{document}

\title{Thermodynamics of Dipolar Chain Systems}

\author{J.~R. Armstrong \and N.~T. Zinner \and D.~V. Fedorov \and A.~S. Jensen }
\institute{ Department of Physics and Astronomy, Aarhus University, 
DK-8000 Aarhus C, Denmark}

\date{\today}
\maketitle
\begin{abstract}
The thermodynamics of a quantum system of layers containing perpendicularly oriented 
dipolar molecules is studied within an oscillator approximation for both
bosonic and fermionic species. The system is assumed to be built from chains 
with one molecule in each layer. We consider the
effects of the intralayer repulsion and quantum statistical requirements in 
systems with more than one chain. Specifically, we
consider the case of two chains and solve the problem analytically within the 
harmonic Hamiltonian approach which is accurate
for large dipole moments. The case of three chains is calculated numerically.
Our findings indicate that thermodynamic observables, such as the heat capacity, 
can be used to probe the signatures of the intralayer interaction between 
chains. This should be relevant for near future experiments on polar molecules
with strong dipole moments.
\end{abstract}
\PACS{
{67.85.-d}{ Ultracold gases, trapped gases } \and
{36.20.-r}{ Macromolecules and polymer molecules } \and  
{05.70.-a}{ Thermodynamics } }

\section{Introduction}
Rheological electro- and magnetofluids are very flexible systems where
long-range dipolar interactions lead to self-organized filamentation. 
Since its discovery in ferrofluids by de Gennes and Pincus \cite{degen1970},
this phenomenon has been intensely studied in classical systems \cite{teix2000}.
This is typically done using Monte Carlo simulations of hard- and soft-sphere
dipolar particles \cite{weis1993-1} and comparison can subsequently be made to 
experiments with colloids that display many 
intriguing phases \cite{puntes2001,yet2003,butter2003,klokkenburg2006}. Important 
technological applications can be found in nanostructuring using atomic 
lithography \cite{oberthaler2003} and in colloidal quantum dot photovoltiac devices
\cite{tang2011}. However, as the systems become increasingly smaller one must
start to consider also quantum aspects of the structure and the dynamics.

The explosive experimental development in ultracold atoms has lead to 
a wealth of new possibilities in terms of simulating the properties of 
condensed-matter systems with extensive control over 
both quantum statistics, interactions, and geometry \cite{lewen2007,bloch2008,chin2010}.
Building on this success, a great deal of effort has been put into the 
cooling of Rydberg atoms, permanent magnetic dipole moment atoms, homo- and heteronuclear molecules, and samples 
of near-degenerate polar molecules can now be produced 
\cite{dipoleexp4,dipoleexp5,dipoleexp6,dipoleexp7,dipoleexp8,dipoleexp9,dipoleexp10,dipoleexp11,miranda2011}.
These systems provide access to strong dipole-dipole forces that are long-range
and anisotropic \cite{theoryrev1,theoryrev2}. In particular, in low-dimensional 
geometries it has been shown that the systems are more stable 
towards two-body chemical reaction loss \cite{miranda2011} and 
to many-body instabilities \cite{theoryrev1,theoryrev2}.

Here we focus on a system consisting of polar molecules that are restricted
to move in a stack of two-dimensional planes. This setup resembles
the classical ferrofluids and colloidal systems with the major exception
that for degenerate or near-degenerate ultracold molecules the system is
intrinsically quantum and physics beyond the classical regime can be 
addressed. An external field can be used
to align the dipoles \cite{micheli2007}, and we assume they are always
perpendicular to the layer planes. Theoretically, a host of 
interesting few- \cite{fewbody1,fewbody2,fewbody3,fewbody4,fewbody5,cremon2010,armstrong2010,artem2011c,artem2011-1,artem2011-2,zinner2011a,armstrong2011a}
and many-body \cite{wang2006,manybody7,manybody9,manybody10,manybody15,manybody16,manybody17,manybody18,manybody22,manybody23,manybody24,manybody25,manybody26,manybody27} states have been predicted in 
single- and multilayer systems. For very large dipole moments, a
triangular crystal phase has been predicted \cite{dipcrystal1,dipcrystal2,dipcrystal3} that is 
similar to the Wigner crystal of the electron gas \cite{wigner1934-2}.

In the current investigation, we 
are interested in the so-called dipolar chain liquid proposed in 
Ref.~\cite{wang2006} for bosonic molecules and later 
also for fermionic molecules \cite{santos2010,potter2010}. For
perpendicularly oriented dipoles, the longest chain possible 
(one molecule in each layer) is always the most bound 
structure \cite{armstrong2011a}.
In the case of bosons, it was found that the longest chain will
condense below a critical temperature \cite{wang2006} which
depends slightly on whether internal vibrational modes of 
the chain are included \cite{wang2008a}. This has been confirmed
by a recent Monte Carlo study \cite{barbara2011}, where a 
lattice model was employed. Common to all the studies mentioned
is that they neglect the repulsive intralayer interaction
of the polar molecules. Here we include this effect in a 
parametric approach and considered the implications on 
thermodynamic quantities such as entropy and heat capacity. 
Thermodynamic variables have been addressed experimentally in the 
case of the short-range interacting cold atomic gases. In particular,
density fluctuation measurements can determine compressibility
and heat capacity \cite{ku2011}. Since absorption 
imaging has been demonstrated also for polar molecules \cite{dipoleexp8},
these measurements should be possible for such systems as well.

Below we present results for the thermodynamic behavior of 
two- and three-chain systems for any number of layers. In the 
case of two chains, we determine the partition function analytically.
Three chains requires, however, numerical calculation of the 
degeneracy of many-body states of given total energy. Our formalism 
is based on an effective harmonic approximation for the two-body interactions
to reduce the Hamiltonian to an exactly solvable form. Naturally, 
this requires a careful choice of parameters for the effective
harmonic interactions. The paper is organized as follows. In 
section~\ref{method} we discuss the Hamiltonian and the 
various approximations we make. As a warm-up, we discuss
in section~\ref{single} the case of a single chain, before
moving onto the case of two- and three-chain systems
in section~\ref{multi}. The resulting thermodynamic results
are presented in section~\ref{results}, along with a discussion 
of experimentally relevant systems. In section~\ref{sumcon}
we summarize our approach and results, and give an 
outlook for future investigations.

\begin{figure}
\includegraphics[width=0.7\textwidth]{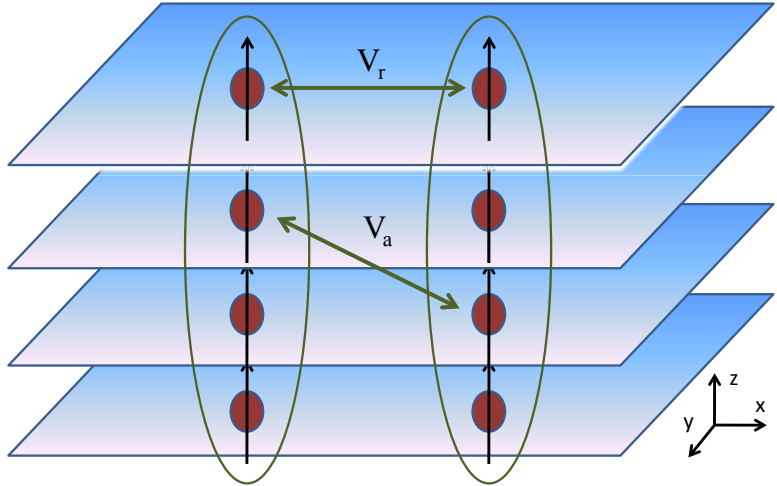}
\centering
\caption{Schematic view of the setup we consider here for four layers with two polar molecules
in each layer. All molecules have dipole moments that are perpendicular to the 
layers as indicated by the black arrows. The chains are indicated by
the ellipsoids. Note, however, that the two molecules in each layer are identical 
particles and that we consider the full Hamiltonian of the system in our study
including the interactions of all molecules in all layers. In the setup shown, the 
full Hamiltonian constitutes an eight-body problem.
The interlayer interaction is denoted $V_a$ (attractive at short distance), while the
intralayer interaction, $V_r$, is purely repulsive. In the lower right-hand corner we 
show the choice of coordinate system.}
\label{schematic}
\end{figure}

\section{Method}\label{method}
The system we consider is shown schematically in Fig.~\ref{schematic}.
In general, we have $N$ molecules of mass $m$ and dipole moment 
$D$ in a layered setup with equidistant layers separated by a 
distance $d$. The layers are assumed to be infinitely thin and
allow no interlayer tunneling. This will be the case for a 
deep optical lattice potential.
We assume that the dipoles are all oriented 
perpendicular to the layers, in which case we have the 
interlayer potential
\begin{eqnarray}
V_a(x,y,n) = D^2 \frac{x^2+y^2 - 2(nd)^2}{(x^2+y^2+(nd)^2)^{5/2}},
\label{va}
\end{eqnarray}
where $x,y$ are relative coordinates of two polar molecules 
that are restricted to move in planes located a distance
$nd$ apart (i.e. adjacent planes have $n=1$ and so forth). 
It is convenient to define the dipolar strength,
$U=\tfrac{mD^2}{\hbar^2d}$, and to measure all energies 
in units of $\tfrac{\hbar^2}{md^2}$, and all temperatures
in units of $\tfrac{\hbar^2}{md^2k_B}$, where $k_B$ is 
Boltzmann's constant. We will also denote the relative radial
distance between two molecules by $r=\sqrt{x^2+y^2}$. The
interaction between two molecules in the same layer is found
by simply setting $n=0$ in Eq.~\eqref{va}, i.e.
\begin{eqnarray}
V_r(r) = D^2 \frac{1}{r^3},
\label{vr}
\end{eqnarray}
a purely repulsive dipolar term.

The Hamiltonian for the $N$-body system is
\begin{equation}
H=-\frac{\hbar^2}{2m} \sum_{k=1}^N
\left(\frac{\partial^2}{\partial x_k^2}+\frac{\partial^2}{\partial y_k^2}\right)
 +\frac{1}{2}\sum_{i\neq k}V_{ik}(x_{ik},y_{ik},n_{ik}) ,
\label{hamil}
\end{equation}
where $(x_{ik},y_{ik})=(x_{i}-x_{k},y_{i}-y_{k})$ is the relative 
coordinate of the pair $(ik)$ with $i,k=1,\ldots,N$, and the factor of 
$\tfrac{1}{2}$ is for double counting. $n_{ik}$
is an index that denotes the distance (in units of $d$) between the layers containing 
molecules $i$ and $k$. For $n_{ik}=0$, $V_{ik}=V_r$ and for $n_{ik}>0$, $V_{ik}=V_a$.
For simplicity, we assume that there is no external confinement in the layer plane itself, but 
this is not essential and can be easily included. Of course, it implies that the 
center of mass motion of the total system must be treated separately. 

To solve the $N$-body problem, we follow the strategy outlined in Ref.~\cite{armstrong2011}
of approximating the interaction terms in the Hamiltonian by quadratic
forms for which the problem can be analytically solved. Using 
simplified models to describe the behavior of self-assembling systems 
that form complex structures has proven very 
successful \cite{ciach2011} in spite of the diverse origin 
of the physical interactions.
For the interlayer
potential in Eq.~\eqref{va}, the harmonic approach has proven very
accurate in reproducing energy, spatial structure, and wave function
\cite{armstrong2010,zinner2011a,armstrong2011a,artem2011d}. The 
real dipole potential is thus replaced by
\begin{eqnarray}
V_{a}^{\textrm{HO}}(r,n) = V_0\left(\frac{r^2}{2(nd)^2}-1\right),
\label{HOpot}
\end{eqnarray}               
where $V_0=V_0(D^2,m,d)$, and the corresponding frequency (which is 
denoted $a$ for attractive) is 
\begin{equation} \label{e40}
\omega_{a}^2=\frac{2V_{0}(D^2,m,d)}{m(nd)^2}.
\end{equation}
This procedure reduces all interlayer
terms to harmonic oscillators that are carefully chosen to reproduce
the two-body energies and the spatial structure of the state 
(more concretely the node of the real potential in Eq.~\eqref{va} 
is reproduced).

It is important to note here that we use the above description for 
the interaction of all pairs of molecules in different
layers. This also includes the interaction of molecules in different
chains and different layers. The spatial structure of the real 
dipole-dipole potential, Eq.~\eqref{va}, has attractive short-range
and repulsive long-range parts. Naively, this could mean that molecules
in different chains would predominantly feel the long-range repulsive
part of the potential. However, we note that this interaction supports
a two-body bound state for any value of the dipole 
strength \cite{armstrong2010,artem2011-1}.
This means that the attractive part of the potential is dominant in 
spite of the spatial profile, and allows one to use a model potential 
that is attractive such as the harmonic oscillator. Of course, the 
strength of the oscillator should be adjusted such that when the 
dipole strength is weak, the harmonic potential is shallow and allows
the wave function to spread out appropriately in space. As shown earlier,
this procedure yields an accurate approximation for the two-body 
wave function that reproduces the energy and root-mean-square radius of the 
real dipole potential for $U\gtrsim 5$ \cite{zinner2011a}. The results
presented here use $U\geq 10$. The interactions between molecules
in different chains and different layers is thus included and will 
compete with the in-plane repulsive interaction described below.

The intralayer repulsion in Eq.~\eqref{vr} can also be approximated by
a harmonic oscillator term. However, it must be of the opposite 
sign compared to the interlayer term in order to describe a 
repulsive force, i.e. the term becomes
\begin{equation}
V_{r}^{\textrm{HO}}(r)=-\frac{1}{4}\omega_{r}^{2}mr^2.
\label{rHO}
\end{equation}
The problem is now to relate $\omega_r$ to the properties of 
the repulsive intralayer dipole-dipole potential. This has been 
discussed at length in Ref.~\cite{armstrong2011a}. 

In the current work, we are interested in the qualitative influence
that the intralayer terms will have when inducing an interaction
between the chains in the system. We therefore use a parametric 
approach and consider different values of $\omega_r$ without 
paying much attention to what the physical value may be. However,
the systems we consider are in the self-bound regime, where the
attractive interlayer terms must overcome the intralayer repulsion.
If we fix $\omega_a$ for given $U$, then there will be a critical 
frequency, $\omega_c$, beyond which the $N$-body harmonic Hamiltonian
will develop complex solutions. This is the signal of instability
toward breaking of the chain system into smaller systems as discussed in 
Ref.~\cite{armstrong2011a}.

Here we consider extremal values;
$\omega_r=0$ and $\omega_r=0.99\omega_c$. These choices will help 
determine what the effects of the chain-chain interaction is 
on thermodynamic observables within the harmonic approximation.
It is then reasonable to expect that the real system will behave
similarly in those limits at least on a qualitative level. The major
point of the current works is thus to investigate how one can determine
the role of chain structures in the system through thermodynamic 
measurements, specifically the changes expected due to quantum statistics
and due to chain-chain interactions.

When we consider the thermodynamics of several chains below it is 
very important to notice that we solve the full $N$-body Hamiltonian
including all molecules in all layers. This means that the chain 
structure indicated on Fig.~\ref{schematic} is not assumed in the 
full Hamiltonian. A different way to approach the problem would 
be to consider the chains as fundamental degrees of freedom. In the 
strong-coupling limit, an effective chain-chain interaction can then
be produced in analogy to the bilayer case \cite{zinner2011a}. 
Scattering of chains could then be calculated and thermodynamics
investigated with respect to this effective potential. However, 
it is less clear how to incorporate the quantum statistical 
identity of the particles in this approach. One could imagine that 
in the limit of very large $U$ where the chains are essentially 
rigid, even length chains will behave as bosons, while odd length
ones will have fermionic nature. This interesting idea is a 
topic for future work. In this investigation we use the full
Hamiltonian and do the symmetrization explicitly in each
layer. This means that we consider {\it all} particles in 
{\it all} layers as equal constituents in the Hamiltonian,
i.e. we do not {\it a priory} consider the chains indicated in
Fig.~\ref{schematic} as the degree of freedom. The use of the 
terms one-, two-, and three-chain systems thus refers explicitly
to the number of particles per layer, and does not imply that
our system consists of one, two, or three rigid chains that interact.

\section{Single Chains}\label{single}
We first consider the case of a single chain, which was 
the focus of Refs.~\cite{wang2006} and \cite{wang2008a}. Since
the geometry is layered and no tunneling is allowed, the
single chain system is equivalent to a system of $N$ 
distinguishable particles. Solving the harmonic 
$N$-body Hamiltonian yields $\mathcal{D}(N-1)$ (where $\mathcal{D}$ is the 
spatial dimension, $\mathcal{D}=2$ in the case
of planes) vibrational 
degrees of freedom since the center of mass separates. 
We denote the mode frequencies by $\omega_i$, $i=1,\ldots,\mathcal{D}(N-1)$. 
Each of these modes will contribute a zero point energy, $\hbar\omega_i$.
This will be part of the overall ground state energy which is 
not our concern here as we focus on the thermal behavior for 
non-zero temperatures and will set the ground state energy
to zero.

The partition function in the canonical ensemble, $Q(N,V,T)$, is 
\begin{equation}
Q(N,V,T)=\sum_l g_l\exp(-\beta E_l),
\label{Qdef}
\end{equation}
where $g_l$ is the degeneracy of the $N$-body state of energy $E_l$, 
and $\beta=1/(k_BT)$ is the inverse temperature.  
For the vibrational spectrum of a single chain,
each mode is non-degenerate and the spectrum is equidistant. The
partition function is therefore a geometric series and we have
\begin{equation}
q_\textrm{vib}^{i}(N,V,T)=\frac{\exp(-\Theta_i/(2T))}{1-\exp(-\Theta_i/T)},
\label{onemode}
\end{equation}
where $\Theta_i=\hbar\omega_i/k_B$. 
Eq.~\eqref{onemode} is the partition function of one mode.  A state with energy, $E_l$, involves 
all the modes, and since the energy 
is in the exponent of Eq.~\eqref{Qdef}, we merely need to take the product 
over all individual modes
\begin{equation}
q_\textrm{sep}(N,V,T)=\prod_i^\Omega\frac{\exp(-\Theta_i/(2T))}{1-\exp(-\Theta_i/T)},
\label{genQsep}
\end{equation}
where $\Omega$ is the total number of vibrational degrees of freedom in 
the system, which as mentioned before is $\Omega=2(N-1)$ when the center of mass
has been factorized. As we will see below, in the case of multiple
chains, the problem will factorize into a term related to single 
chain behavior, $q_\textrm{sep}$, and a term containing the chain-chain
modes.

The center of mass motion can easily be taken into account. In the 
case where there is no external confinement in the layers, we 
simply multiply by the well-known non-interacting translational
partition function
\begin{equation}
q_\textrm{trans}=\left[\frac{mk_BT}{2\pi\hbar^2}\right]^{\mathcal{D}/2}V,
\end{equation}
where $V$ is the volume and $D$ is the dimension.  
The total partition function is $Q=q_\textrm{sep}q_\textrm{trans}$.  
If, however, there is external confinement in terms of an 
in-plane quadratic field, the number of vibrational degrees of freedom 
is simply $\mathcal{D}N$ with $\mathcal{D}$ frequencies that are exactly given by 
the external confinement frequency. The partition function can be
computed simply from Eq.~\eqref{genQsep}. If the system isotropic is
and the center of mass separates, 
then Eq.~\eqref{genQsep} can be simplified to
\begin{equation}
q_{sep}(N,V,T)=\left(\prod_i^{\Omega}\frac{\exp(-\Theta_i/(2T))}{1-\exp(-\Theta_i/T)}\right)^\mathcal{D}.
\label{Qiso}
\end{equation}

Notice that our description of the single chain includes {\it all} possible
internal vibrational modes similar to the approach of Ref.~\cite{wang2008a}.
However, we fix the potential to reproduce physical two-body features
of the interlayer dipole-dipole interaction, whereas Ref.~\cite{wang2008a}
uses an expansion of the dipolar potential around the origin to second
order. Both approaches yield accurate results for energies of longer
chains when compared to exact numerics \cite{artem2011d}.

\section{Two and Three Chains}\label{multi}
In order to explore the effect of the intralayer repulsion, we need to 
extend to at least two molecules per layer. The minimal model would 
thus be a two-chain system. Ideally we would like to consider
the thermodynamics for any number of chains. However, the requirements
of quantum statistics makes this an extremely complicated problem to 
solve due to the computational effort involved in determining the 
degeneracy of the many-body states. Fortunately, systems of ultracold
molecules often have very low density, and in this case we expect the 
dynamics of a few chains to capture the leading correlations
in the system. The reasons for considering both two {\it and} three
chain systems is that the dipolar interaction is expected to 
induced a triangular crystal in the limit of very strong dipole 
moment \cite{dipcrystal1,dipcrystal2,dipcrystal3}. A precursor of this transition should be 
present in a three chain model.
For high density, one would need a different starting point, 
typically mean-field theory. 
We concentrate on this low-density limit from now on and calculate the
properties of two- and three-chain systems.

When dealing with identical particles, one must take into account the 
appropriate symmetry requirements of the wave function of those particles.  
For identical fermions, the wave function should be antisymmetric, 
and for identical bosons, symmetric. The ground state of the harmonic 
solution 
is a symmetric Gaussian wave function, which could thus be a good approximation
to a Bose system. Once we consider excited states for use in the partition
function, the problem is much tougher, especially for many 
particles, as most excitations will, at first glance, appear neither 
symmetric nor antisymmetric. However, as we discuss below, in the case
of two chains, the required symmetrization can be done analytically. For
three chains one must resort to a numerical determination of the 
partition function \cite{jeremy2011}.

\subsection{Two Chains}
For two chains (as shown in Fig.~\ref{schematic}),
molecules in different layers can be treated as
distinguishable particles, while quantum statistics are important 
for molecules in the same layer.  
The total partition function can be written as a
product of the separable single-chain term of Eq.~\eqref{genQsep} 
and an intralayer quantum statistically restricted term
\begin{equation}
Q=q_\textrm{sep}q_{f/b},
\label{Qtotal}
\end{equation}
where $q_{f/b}$ is the boson/fermion intralayer part.  

From diagonalization of the $N$-body Hamiltonian, we obtain
$2W-1$ \cite{armstrong2011} frequencies, where $W$ is the number of layers, and the 
center of mass has been separated. Out of these frequencies, $W-1$ 
refer to the single chain modes. The remaining $W$ frequencies describe 
the intralayer repulsion \cite{armstrong2011a}.  
Each of these $W$ frequencies refer to a particular mode
in the diagonalizing coordinates. When one writes these 
new coordinates in terms of the original laboratory 
coordinates \cite{armstrong2011} 
introduced above, one sees that they are in 
fact antisymmetric in the coordinates of the identical 
molecules in each layer, i.e. they correspond to 
traditional relative coordinates, $\bm r_{1,k}-\bm r_{2,k}$, 
where $k=1,\ldots,W$ is a layer index. This immediately 
implies that exciting any one of these $W$ intralayer
modes by an even number of quanta will produce a symmetric
state under exchange of two particles in the layer, while
an odd number of quanta produces an antisymmetric state.
In turn, this can be used to construct the excitations that 
are allowed for given quantum statistics.

\subsubsection{Bosons}
Bosons require a symmetric wave function under interchange in each plane.
In two dimensions, the relative in-plane wave function is characterized
by two non-negative quantum number, $n_x$ and $n_y$, an independent harmonic oscillator 
for each spatial coordinate. It is then clear that for 
bosonic molecules, a symmetric wave function can be achived if either 
both $n_x$ and $n_y$ are even or both are odd. In this case we therefore
have $n_x+n_y=2j$ for integer $j\geq 0$. For given $j$, the degeneracy 
of this state is $2j+1$ (remember that the state $n_x=n_y=j$ is non-degenerate).
The partition function for bosons now becomes
\begin{equation}
q_b=\prod_k^W\left(\sum_{j=0}(2j+1)\exp(-2j\Theta_k/T)\right),
\end{equation}
where the product runs over all layers, $k=1,\ldots,W$, while the sum runs 
to infinity.  This expression can be written 
\begin{eqnarray}
q_b&=&\prod_k^W\left[\frac{1+\exp(-2\Theta_k/T)}{[1-\exp(-2\Theta_k/T)]^2}\right].
\label{qdeg}
\end{eqnarray}

\subsubsection{Fermions}
For fermions, we require antisymmetry with respect to particle exchange.  
This is achieved by enforcing that the total excitation in each direction is odd, 
i.e. $n_{x}+n_{y}=2j+1$ for $j\geq 0$. The degeneracy can be counted in the 
same way as for bosons, but the requirement of an odd sum now yields $2j$. 
We have
\begin{equation}
q_f=\prod_k^W\left(\sum_{j=0}2j\exp(-(2j-1)\Theta_k/T)\right),
\end{equation}
which becomes
\begin{eqnarray}
q_f&=&\prod_k^{W}\left[\frac{2\exp(-2\Theta_k/T)}{[1-\exp(-2\Theta_k/T)]^2}\right].
\label{qf1st}
\end{eqnarray}

\subsubsection{Extensions}
The symmetry considerations above can be straightforwardly extended to three
dimensions. In addition, an in-plane external confinement potential 
given by a harmonic oscillator does not change anything. The center
of mass mode will be symmetric in all coordinates so it merely
modifies the separable part of the partition function as discussed 
above. Thus, it does not modify the symmetry properties.

In the oscillator formalism, a constant 
can be added to the Hamiltonian in order to achieve some desired 
property of the energy of the system. Here we are interested in 
the behavior of the system as it is thermally excited, and 
not in the ground state energy. We therefore choose to set the 
zero of energy at the ground state level, which is of course
non-zero in absolute terms due to zero-point contributions of 
all oscillator terms.

\begin{figure*}
\includegraphics[width=0.98\textwidth]{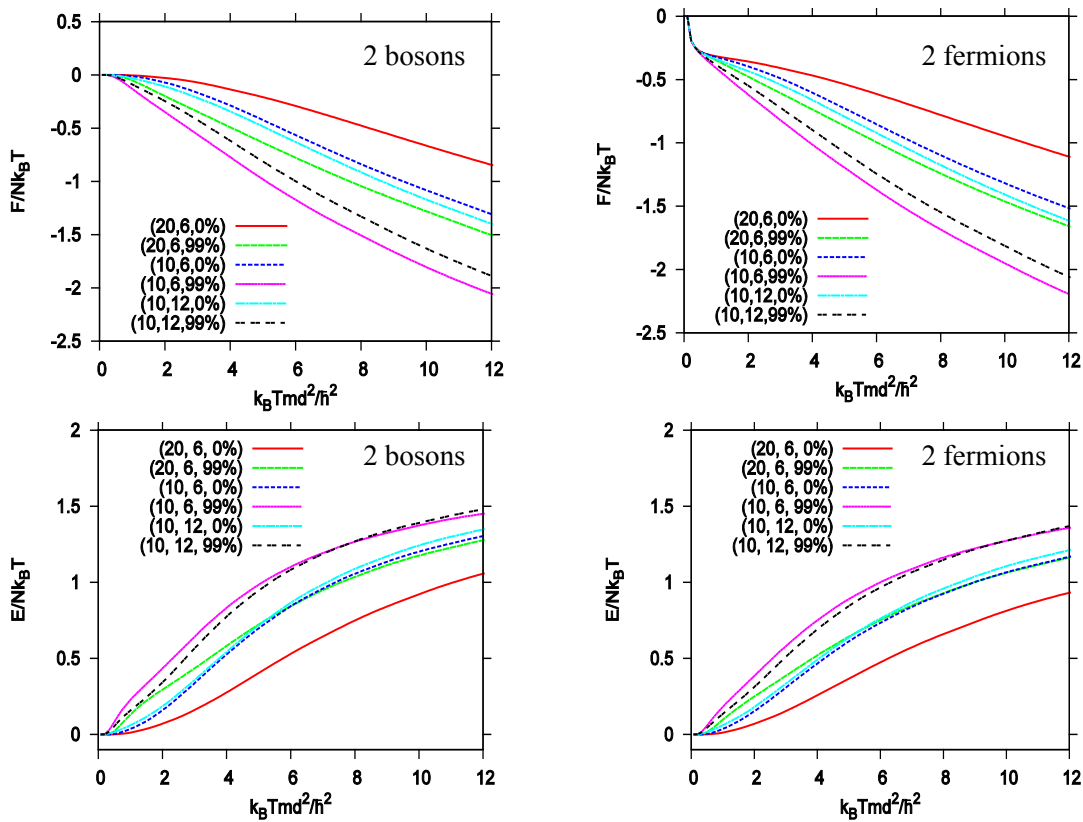}
\caption{Helmholtz free energies, $F$ (upper row), and energies, 
$E$ (lower row), divided by $Nk_BT$ for two bosonic (left column) 
and two fermionic (right column) chains. The different lines 
correspond to different values of the dipole strength, $U$, 
the number of layers, $W$, and the magnitude of the repulsive 
intralayer frequency, $\omega_r$, measured in percentage of the 
critical frequency at which the system breaks apart. The 
nomenclature is $(U,W,\tfrac{\omega_r}{\omega_c})$.}
\label{bfAE}
\end{figure*}

\subsubsection{Thermodynamic quantities}
Having obtained the partition function, $Q$, analytically in the 
case of two chains, thermodynamic observables can be calculated.  
In the canonical ensemble, the most straightforward is the 
Helmholtz free energy, $F$, which 
is the characteristic thermodynamic function of the canonical partition function.
In the case of distinguishable particles, we have
\begin{eqnarray}
F&=&-k_BT\ln Q\nonumber \\
&=&k_BT\sum_j^\Omega\left(\frac{\Theta_j}{2T}+\ln(1-\exp(-\Theta_j/T))\right),
\end{eqnarray}
where the sum is taken over all vibrational degrees of freedom in the system.  
If the system is isotropic, then the free energy is
\begin{equation}
F=\mathcal{D}k_BT\sum_j^{W-1}\left(\frac{\Theta_j}{2T}+\ln(1-\exp(-\Theta_j/T))\right).
\end{equation}
The energy, $E=k_BT^2\tfrac{\partial\ln Q}{\partial T}$, 
the constant volume heat capacity, $C_V=\left(\tfrac{\partial E}{\partial T}\right)_V$, and
the entropy, $S=k_B\ln Q+k_BT\tfrac{\partial\ln Q}{\partial T} $, is easily obtained as well.
The total partition function for indistinguishable particles 
is $Q=q_\textrm{sep}q_{f/b}$. The partition function can thus be separated, 
and the 
above formulae can be applied, with the appropriate limits on the sums.

\begin{figure*}
\includegraphics[width=0.98\textwidth]{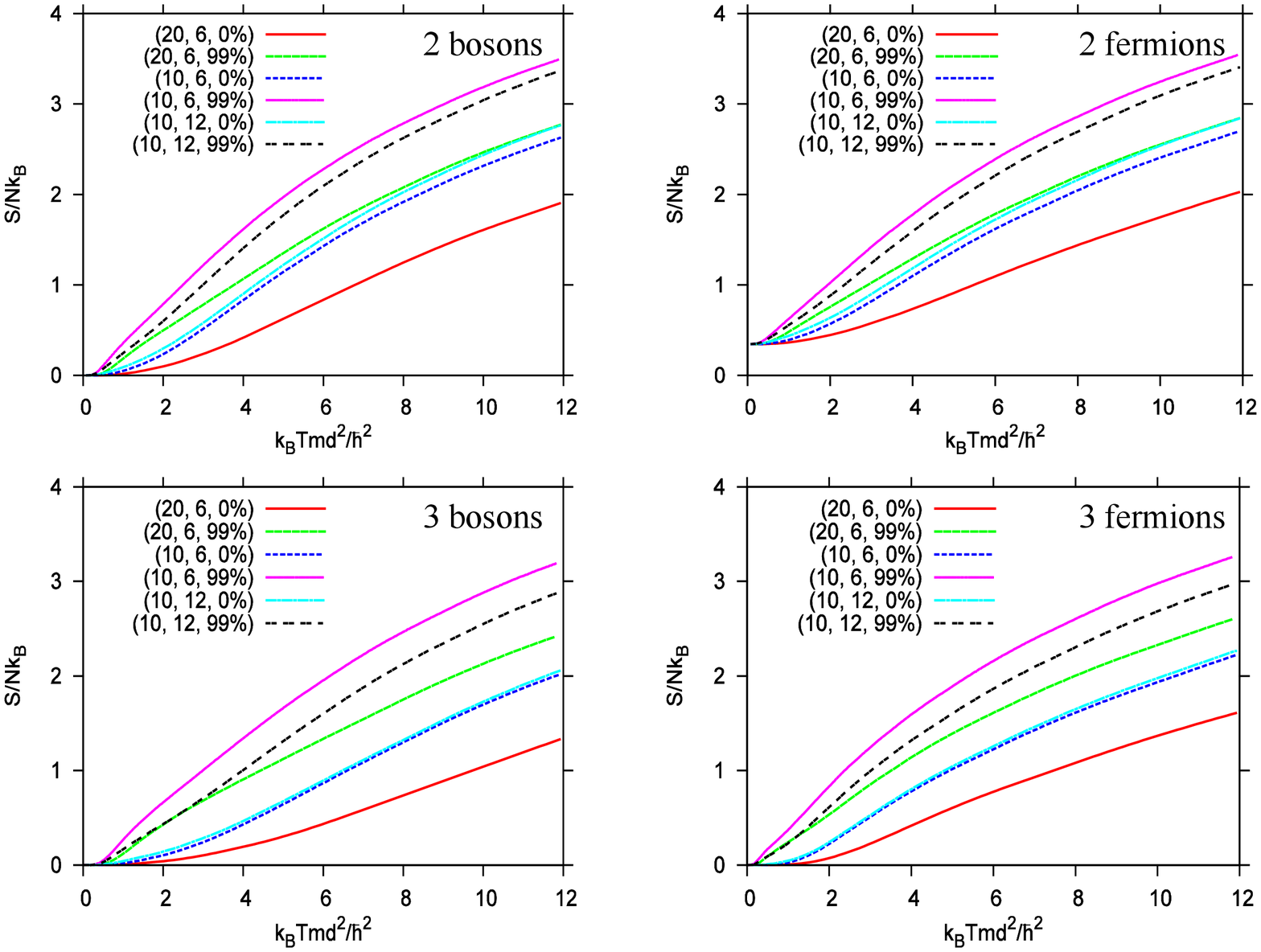}
\caption{Entropies, $S$, divided by $Nk_B$ for systems of two (upper row) and 
three (lower row) chains of bosonic (left column) and fermionic (column) 
molecules. The nomenclature of the legends is the same as in Fig.~\ref{bfAE}.}
\label{bfS}
\end{figure*}

\subsection{Three Chains}
The analytical results above for the partition function in the 
two-chain case can unfortunately not be extended to a larger number
of chains. In this case one needs to return to the general partition
function Eq.~\eqref{Qdef} and compute this numerically. As demonstrated
in Ref.~\cite{jeremy2011}, this can be done quite effectively
in the harmonic approximation for both bosonic and fermionic 
particles. The three-chain case has been computed numerically in 
order to obtain $q_{f/b}$ from Eq.~\eqref{Qtotal} that goes along 
with $q_\textrm{sep}$ to obtain Eq.~\eqref{Qdef}.

\section{Results}\label{results}
We now present the results for both two and three chain systems for different
values $U=10$ and 20, for $\omega_r=0$ and $\omega_r=0.99\omega_c$, and for
layer numbers $W=6$ and 12. These values of $U$ are both in the regime where
the harmonic approximation to $V_a$ of Eq.~\eqref{va} is extremely accurate.
We first discuss Helmholtz free energy and energy for two chains, omitting the 
case of three chains which is very similar. Next we present the entropies for
two and three chains, and finally we show the heat capacities. We specialize to
also discuss the distinct low temperature behavior of the heat capacity where
quantum statistics is most pronounced.

\subsection{Energy}
In Fig.~\ref{bfAE}, we plot the Helmholtz free energy (upper row) 
and energy (lower row) for bosons (left column) and fermions (right column)
for two-chain systems. An immediate striking difference between 
the two upper plots, is the behavior at zero temperature where bosons
are smooth, whereas fermions have a kink. This is a consequence 
of the ground state degeneracy of the fermionic system and appears through
the entropy term in $F=E-TS$. We will discuss this in more detail 
when we present entropies below. At slightly higher temperatures, $F$ 
agrees quite well for bosons and fermions, with fermions having 
lower overall values in all cases. In the case of $\omega_r=0.99\omega_c$, 
the curves are considerably below the $\omega_r=0$ case. This is a 
consequence of the fact that the mode frequencies are lowered
as $\omega_r$ grows \cite{armstrong2011a}, allowing easier thermal 
excitation of the system. We also note that changing the layer 
number from 6 to 12 induces only slight quantitative changes. 
The energies in the lower row of Fig.~\ref{bfAE} show the same 
tendencies as $F$, and their close similarity attests to the fact 
that the entropy term is causing the differences in $F$ at low temperature.

\subsection{Entropy}
The entropies of boson (left column) and fermion (right column) 
two (upper row) and three (lower row) chain systems are shown in 
Fig.~\ref{bfS}.
For two fermionic chains we see that the entropy does {\it not} 
go to zero at zero temperature. This is caused by the fact that 
antisymmetry 
requires at least one quantum of excitation. For two-dimensional 
oscillators the first excited state is degenerate and this implies 
finite entropy at zero temperature. Due to the factor of 2 inside
the product in Eq.~\eqref{qf1st}, the zero 
temperature contribution is $k_BW\ln 2$, so the plot approaches $\ln 2/2$ for 
many layers (recall that $N$ is the {\it total} number of particles,
while $W$ is the number of layers, i.e. $W/N=1/2$). For three chains, 
the ground state is non-degenerate in two dimensions (the oscillator
has a non-degenerate ground state and a twice degenerate first excited 
state). Thus the entropy goes to zero at zero temperature for three 
chains irrespective of quantum statistics.

We notice that the entropy starts to increase at larger temperatures 
for bosons than for fermions, implying that the activation energy gap 
is larger for bosons. This will become more pronounced when we consider heat 
capacities below and we postpone the discussion. In general, 
we see that for larger $\omega_r$, the entropy in general grows for 
all temperatures as the modes go down in energy as discussed above. 
Again we observe that the quantum statistics does not seem to play
much role as the left and right columns in Fig.~\ref{bfS} are very similar.

\begin{figure*}
\includegraphics[width=0.98\textwidth]{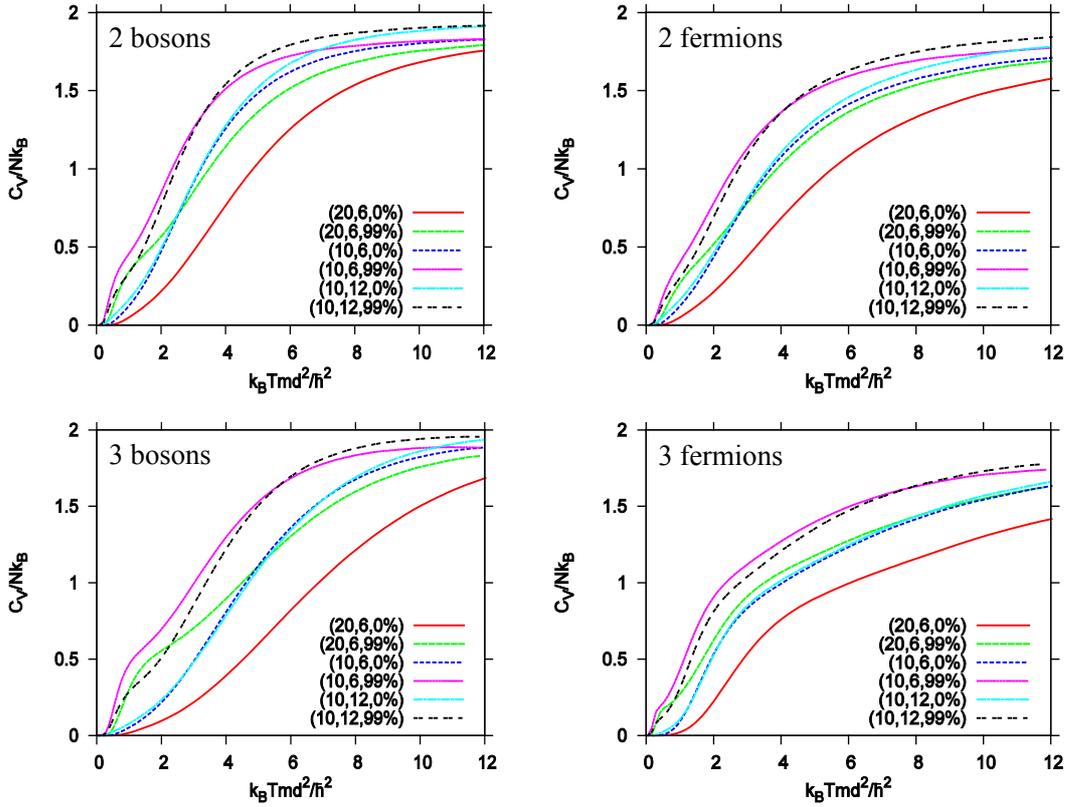}
\caption{Heat capacity, $C_V$, divided by $Nk_B$ for systems of two 
(upper row) and three (lower row) chains of bosonic (left column) and fermionic (column) 
molecules. The nomenclature of the legends is the same as in Fig.~\ref{bfAE}.}
\label{bfCV}
\end{figure*}

\subsection{Heat Capacity}
The derivatives of the 
heat capacity makes it very sensitive to changes in behavior due to various modes that become
activated in the system. This can be seen for instance in second order phase 
transitions where the heat capacity will become singular at the critical temperature. As we
discussed in the introduction, the heat capacity is now a measurable quantity in ultracold 
atomic gases \cite{ku2011}. We therefore anticipate that similar measurements should 
be possible also for experiments with polar molecules in the near future. Another good 
observable would be the compressibility of the system. The detailed 
balance between attractive and repulsive interactions should be visible there as well.

In Fig.~\ref{bfCV} we show a plot similar to Fig.~\ref{bfS} for the heat capacities for
the same temperature ranges employed in Figs.~\ref{bfAE} and \ref{bfS}. Overall, we 
note the high temperature behavior of all cases which approaches the equipartition 
value of $2Nk_B$ characteristic of two-dimensional oscillators. Another general feature
is that for larger $U$, the heat capacity decreases due to the larger cost of 
excitation in the system. When we compare fermions
to bosons, we find that fermions in general approach the limit slightly slower for 
both two and three chain systems. There is a clear 'shoulder' at low temperature 
which seems to be more pronounced for bosons, but appears at lower temperatures for
the fermions. Also, going from two to three chains seems to make the 'shoulder' more
pronounced. This feature is clearly associated with the larger value of $\omega_r$, 
and also becomes more clear with increasing $W$. This implies that measurements
of the heat capacity can potentially yield information about the role that 
chain-chain interaction plays in a dipolar chains liquid.

\begin{figure*}
\includegraphics[width=0.98\textwidth]{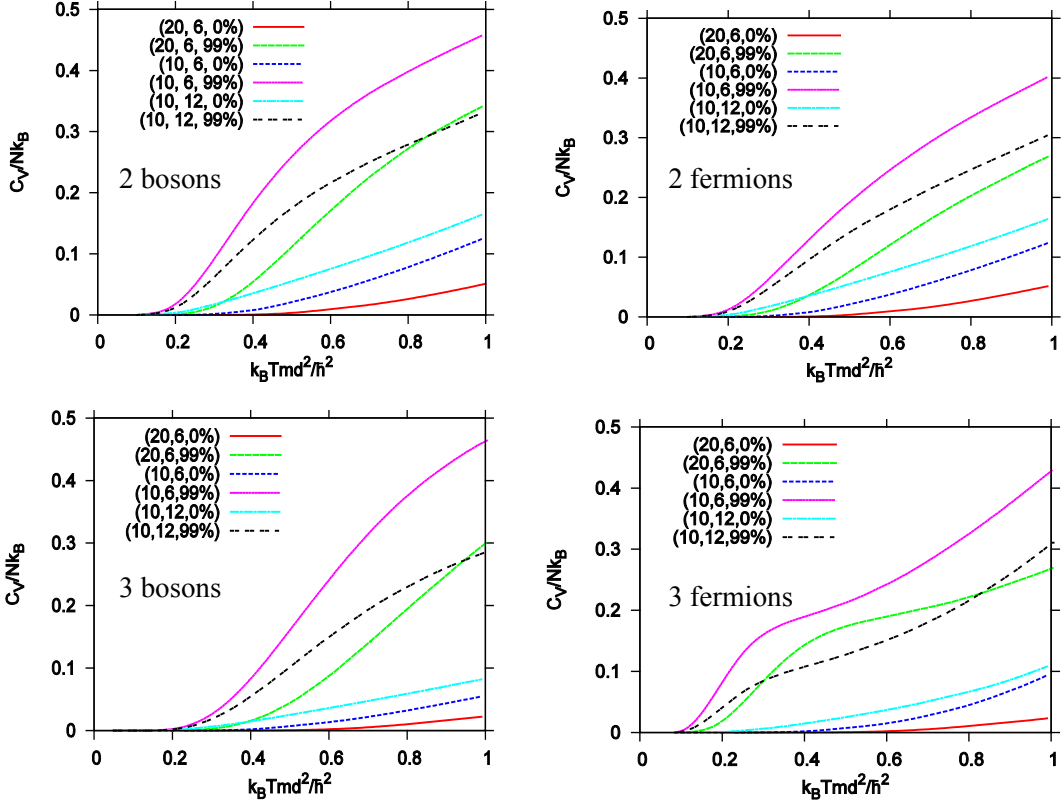}
\caption{Same as Fig.~\ref{bfCV} but for a smaller range of temperatures around zero.}
\label{bfCVzoom}
\end{figure*}

In order to further investigate the interesting features at low 
temperature in the heat capacities, we show a plot where the 
range of temperature is zoomed in around the low temperature
region in Fig.~\ref{bfCVzoom}. Here we see that there is 
striking differences between fermions and bosons, and 
between the two and three chain systems. Not unexpectedly, 
the effects of intralayer repulsion is most pronounced at 
low temperatures. The clear 'shoulder' seen for large 
$\omega_r$ and three fermions is the one we see at 
higher temperatures for bosons in Fig.~\ref{bfCV}. This can 
be understood from the fact that bosons have larger activation
energies due to a gap in the low energy spectrum. For 
bosons one cannot generate a totally symmetric state
by promoting one molecule from the ground
to the first excited state, since the wave function would 
then necessarily become antisymmetric. Thus, the gap 
for bosons is twice the level spacing in the low-energy
spectrum, whereas for fermions this is allowed and the gap
is just the level spacing. This is true for the three 
chain systems as seen clearly in the lower row of 
Fig.~\ref{bfCVzoom}. For two chains, we find that the gap
is always the same (twice the level spacing) for both
kinds of molecules, and the stronger increase of $C_V$
for bosons at low temperature is due to a larger density 
of states \cite{jeremy2011}.
Note also the difference for 
three fermions for the two different $\omega_r$ values, 
with the larger one giving a much smaller activation energy 
due to the lowering of the mode frequencies.

It is interesting to note that the heat capacity 
grows with $W$ when $\omega_r=0$, while the opposite is 
true for large $\omega_r$. This is connected to the fact that
in the former case, the longer chains have more (degenerate)
modes to thermally excite, whereas in the latter case the 
repulsion is pushing all modes down toward zero but slower
for longer chains where more attractive terms are present. 
This means that measurements of heat capacities for systems
of different $W$ can help determine whether the repulsive
intralayer interaction is important or not for given system
density.

\subsection{Relation to Current Experiments}
For comparison to experiment it is convenient to recall our
units for dipolar strength and for temperature. Here we 
use that 1 Debye equals 
$3.336\cdot 10^{-30}\,\textrm{C}\,\textrm{m}$ in SI units. 
Also we must remember that the dipole-dipole interaction in 
SI units has a factor of $\tfrac{1}{4\pi\epsilon_0}$, where
$\epsilon_0$ is the permittivity of vacuum.
The dimensionless dipolar strength can be 
written as
\begin{eqnarray}
U=0.015\, \left(\frac{m}{u}\right)\left(\frac{D}{1\,\textrm{Debye}}\right)^2
\left(\frac{1\,\mu\textrm{m}}{d}\right),
\end{eqnarray}
where $u$ is the atomic mass unit. Similarly, the unit of temperature
that we use can be expressed in the form
\begin{eqnarray}
\frac{\hbar^2}{md^2}\frac{1}{k_B}=485\,\left(\frac{u}{m}\right)
\left(\frac{1\,\mu\textrm{m}}{d}\right)^2\,\textrm{nK}.
\end{eqnarray}

The experiments most relevant to our setup are the ones
being conducted at the moment at JILA \cite{miranda2011}. 
The rotational and vibrational ground state molecules used 
are $^{40}$K$^{87}$Rb which has a maximum dipole moment
of $D=0.566$ Debye. The experiment reported in 
Ref.~\cite{miranda2011} has studied chemical properties 
of a sample confined to a quasi-two-dimensional stack of 
about 23 layers and found profound effects due to the 
optical lattice potential on the loss rates. The dipoles
are perpendicular to the planes and the applied field has
a magnitude that induces a dipole moment of $D=0.158$ Debye.
With a lattice spacing of 532 nm, this amounts to $U=1.15$
for maximum dipole moment, while for the experiment cited
it is roughly $U=0.09$. Our results are based on the 
strong-coupling harmonic approximation, and $U$ must 
therefore be large. The JILA experiments are therefore
not in the regime of validity of our approach. 

To get a stronger dipole moment, NaK molecules
consisting of a Sodium and a Potassium atom can be used.
They are predicted to have a maximum moment of 
about $D=2.7$ Debye \cite{aymar2005}. This has the 
advantage that several isotopes of Potassium are 
available for ultracold atomic experiments of 
both bosonic and fermionic character.
Assuming that $d=532$ nm, one could in principle 
reach values of $U\sim 13$ which is within the region
where we expect our approximations to hold.
Furthermore, the range of temperature shown in
Figs.~\ref{bfAE}-\ref{bfCV} above corresponds to 27 nK in this case. We thus 
expect that intralayer effects could be studied for system 
temperatures below about 270 nK. Reaching this scale should 
not be difficult experimentally.

\section{Summary and Outlook}\label{sumcon}
The physics of dipolar chain systems in multilayered 
geometries holds great potential for realization of 
exotic few- and many-body physics that is hard to 
access in other setups. Here we have considered the 
case of chains of dipolar molecules with dipole
moment oriented perpendicular to the layer planes 
and studied the thermodynamics for both 
bosonic and fermionic molecules. In order to 
render the problem tractable, we approximate the
real Hamiltonian by a harmonic form with suitably
chosen parameters that reproduce realistic two-body
properties of dipolar systems.

In contrast to previous studies of the dipolar 
chain liquid, we include the effect of interactions
between the chains by introducing a repulsive 
harmonic term that parametrizes this effect. We have
also explicitly taken care of the quantum statistical 
effects of having more than one molecule per layer. 
In the simplest non-trivial case, there are two chains
in the system and for that we can determine the partition
function and in turn all thermodynamic properties
analytically within the harmonic approach. In the 
case of three chains, the degeneracies of the states
could no longer be determined analytically and we 
resort to numerics. 

Our findings indicate that heat capacity measurements
for the system can probe the relative influence of 
excitation within single chains and the inter-chain 
dynamics. While we have only considered the two and 
three chain cases here, for a low density system
we expect these to be the leading 
correlations in the system. A viable way to study 
our predictions could be through measurements at 
different densities. Starting from very low 
density, we would expect that the single chain
modes are more important and will dominate the 
thermodynamics. However, as the density is 
increased, one should start to see effects of the 
inter-chain modes. In the case where the 
repulsive intralayer interaction is very strong
we find a softening of the modes of the system and 
thus an increase in the heat capacity that could 
be observable. Another observable that we expect 
will be strongly influenced by inter-chain dynamics
is the compressibility. In the limit of very large 
density, we expect our approach to become less 
accurate and mean-field theory is probably
a better starting point.

The melting of classical crystals including those with dipolar 
interactions have been studied several decades ago \cite{kalia1981-1}, and 
recently the corresponding quantum system in the case of 
a single layer with perpendicular dipoles was addressed \cite{dipcrystal1,dipcrystal2,dipcrystal3}.
The interest is focused on the ability of dipolar systems to form
a crystal akin to the Wigner crystal of the two-dimensional electron
gas \cite{wigner1934-2}. The results presented here suggest that in the 
multilayer case there can be an interesting competition between 
interlayer attractive interactions and the in-plane repulsion that 
could modify the crystallization. This is an interesting topic for
future studies.

Another question that arises for a system consisting of dipolar chains
is whether one can develop an effective description that considers the 
chains themselves as the basic constituents. This is conceivable in 
the limit of low temperatures where the occupancy of higher modes
is small. The quantum statistics of the chain would then depend on 
whether it has an even (bosonic) or odd (fermionic) number of molecules.
At first glance, the former would constitute a (repulsive) Bose 
gas, while the latter
would correspond to a Fermi liquid-like state.
The effective interaction between these entities could help
discern potential instabilities in the system, such as the roton
instability for bosons or the density-wave instability for fermions.
In the limit where the number of layers is large the two types
of instability should merge. It would be interesting to try
to come up with an analytical expression for the dynamics in this
limit. Here the harmonic approximation could be very useful.


\begin{thebibliography}{99}
\bibitem{degen1970} de Gennes P.G. and Pincus P.A.: Pair correlations in a ferromagnetic colloid. Phys. Kondens. Mater. {\bf 11}, 189 (1970)

\bibitem{teix2000} Teixeira P.I.C., Tavares J.M., Telo da Gama M.M.: The effect of dipolar forces on the structure and thermodynamics of classical fluids. J. Phys. Condens. Matter {\bf 12}, R411 (2000)

\bibitem{weis1993-1} Weis J.J. and D. Levesque D.: Chain formation in low density dipolar hard spheres: A Monte Carlo study. Phys. Rev. Lett. {\bf 71}, 2729 (1993)

\bibitem{puntes2001} Puntes V.F., Krishnan K.M., Alivisatos A.P.: Colloidal nanocrystal shape and size control: The case of cobalt. Science {\bf 291}, 2115 (2001)
\bibitem{yet2003} Yethiraj A. and van Blaaderen A.: A colloidal model system with an interaction tunable from hard sphere to soft and dipolar. Nature {\bf 421}, 513 (2003)
\bibitem{butter2003} Butter K. {\it et al.}: Direct observation of dipolar chains in iron ferrofluids by cryogenic electron microscopy. Nature Materials {\bf 2}, 88 (2003)
\bibitem{klokkenburg2006} Klokkenburg M., Dullens R.P.A., Kegel W.K., Erne B.H., Philipse A.P.: Quantitative Real-Space Analysis of Self-Assembled Structures of Magnetic Dipolar Colloids. Phys. Rev. Lett. {\bf 96}, 037203 (2006)

\bibitem{oberthaler2003} Oberthaler M.K. and Pfau T.: One-, two- and three-dimensional nanostructures with atom lithography. J. Phys.: Condens. Matter {\bf 15}, R233 (2003)

\bibitem{tang2011} Tang J. {\it et al.}: Colloidal-quantum-dot photovoltaics using atomic-ligand passivation. Nature Materials {\bf 10}, 765 (2011)

\bibitem{lewen2007} Lewenstein M. {\it et al.}: Ultracold atomic gases in optical lattices: mimicking condensed matter physics and beyond. Adv. Phys. {\bf 56}, 243 (2007)
\bibitem{bloch2008} Bloch I., Dalibard J., Zwerger W.: Many-Body Physics with Ultracold Gases. Rev. Mod. Phys. {\bf 80}, 885 (2008)
\bibitem{chin2010} Chin C., Grimm R., Julienne P.S., Tiesinga E.: Feshbach Resonances in Ultracold Gases. Rev. Mod. Phys. {\bf 82}, 1225 (2010)

\bibitem{dipoleexp4} Ospelkaus, S. {\it et al.}: Efficient state transfer in an ultracold dense gas of heteronuclear molecules. Nature Phys. {\bf 4}, 622 (2008)
\bibitem{dipoleexp5} Ni, K.-K. {\it et al.}: A High Phase-Space-Density Gas of Polar Molecules. Science {\bf 322}, 231 (2008)
\bibitem{dipoleexp6} Deiglmayr, J. {\it et al.}: Formation of Ultracold Polar Molecules in the Rovibrational Ground State. Phys. Rev. Lett. {\bf 101}, 133004 (2008)
\bibitem{dipoleexp7} Lang, F., Winkler, K., Strauss, C., Grimm, R., Hecker Denschlag, J.: Ultracold Triplet Molecules in the Rovibrational Ground State. Phys. Rev. Lett. {\bf 101}, 133005 (2008)


\bibitem{dipoleexp8} Wang D. {\it et al.}: Direct absorption imaging of ultracold polar molecules. Phys. Rev. A {\bf 81}, 061404(R) (2010)
\bibitem{dipoleexp9} Ni, K.-K. {\it et al.}: Dipolar collisions of polar molecules in the quantum regime. Nature {\bf 464} 1324, (2010)
\bibitem{dipoleexp10} Ospelkaus, S. {\it et al.}: Quantum-State Controlled Chemical Reactions of Ultracold Potassium-Rubidium Molecules. Science {\bf 327} 853, (2010)
\bibitem{dipoleexp11} Sawyer B.C. {\it et al.}: Cold heteromolecular dipolar collisions. Phys. Chem. Chem. Phys. {\bf 13}, 19059 (2011)


\bibitem{miranda2011} de Miranda, M.H.G. {\it et al.}: Controlling the quantum stereodynamics of ultracold bimolecular reactions. Nature Phys. {\bf 7}, 502 (2011)
\bibitem{theoryrev1} Baranov, M.~A.: Theoretical progress in many-body physics with ultracold dipolar gases. Phys. Rep. {\bf 464}, 71 (2008)
\bibitem{theoryrev2} Lahaye, T., Menotti, C., Santos, L., Lewenstein, M., Pfau, T.: The physics of dipolar bosonic quantum gases. Rep. Prog. Phys. {\bf 72}, 126401 (2009)


\bibitem{micheli2007} Micheli, A., Pupillo, G., B{\"u}chler, H.P., Zoller, P.: Cold polar molecules in two-dimensional traps: Tailoring interactions with external fields for novel quantum phases. Phys. Rev. A {\bf 76}, 043604 (2007)


\bibitem{fewbody1} Shih, S.M., Wang, D.W.: Pseudopotential of an interaction with a power-law decay in two-dimensional systems. Phys. Rev. A {\bf 79}, 065603 (2009)
\bibitem{fewbody2} Klawunn, M., Pikovski, A., Santos, L.: Two-dimensional scattering and bound states of polar molecules in bilayers. Phys. Rev. A {\bf 82}, 044701 (2010)
\bibitem{fewbody3} Fedorov, D.V., Armstrong, J.R., Zinner, N.T., Jensen, A.S.: Weakly bound states of polar molecules in bilayers. Few-body Syst. {\bf 50}, 417 (2011)
\bibitem{fewbody4} Wunsch, B., Zinner, N.T., Mekhov, I.B., Huang, S.J., Wang, D.W., Demler, E.: Few-body bound states in dipolar gases and their detection. Phys. Rev. Lett. {\bf 107}, 073201 (2011)
\bibitem{fewbody5} Zinner, N.T., Wunsch, B., Mekhov, I.B., Huang, S.J., Wang, D.W., Demler, E.: Few-Body Bound Complexes in One-dimensional Dipolar Gases and Non-Destructive Optical Detection. Phys. Rev. A {\bf 84}, 063606 (2011) 


\bibitem{cremon2010} Cremon, J.C., Bruun, G.M., Reimann, S.M.: Tunable Wigner States with Dipolar Atoms and Molecules. Phys. Rev. Lett. {\bf 105}, 255301 (2010)
\bibitem{armstrong2010} Armstrong, J.R., Zinner, N.T., Fedorov, D.V., Jensen, A.S.: Bound states and universality in layers of cold polar molecules. Europhys. Lett. {\bf 91}, 16001 (2010)
\bibitem{artem2011c} Volosniev, A.~G., Fedorov, D.~V., Jensen, A.~S., Zinner, N.~T.: Few-body bound state stability of dipolar molecules in two dimensions. Phys. Rev. A {\bf 85}, 023609 (2012)  
\bibitem{artem2011-1} Volosniev, A.G. {\it et al.}: Bound dimers in bilayers of cold polar molecules. J. Phys. B {\bf 44}, 125301 (2011)
\bibitem{artem2011-2} Volosniev, A.G., Fedorov, D.V., Jensen, A.S., Zinner, N.T.: Model Independence in Two Dimensions and Polarized Cold Dipolar Molecules. Phys. Rev. Lett. {\bf 106}, 250401 (2011)
\bibitem{zinner2011a} Zinner, N.T., Armstrong, J.R., Volosniev, A.G., Fedorov, D.V., Jensen, A.S.: Dimers, Effective Interactions, and Pauli Blocking Effects in a Bilayer of Cold Fermionic Polar Molecules. Few-Body Syst. in press, arXiv:1105.6264
\bibitem{armstrong2011a} Armstrong, J.R., Zinner, N.T., Fedorov, D.V., Jensen, A.S.: Layers of Cold Dipolar Molecules in the Harmonic Approximation. Eur. Phys. J. D {\bf 66}, 85 (2012)

\bibitem{wang2006} Wang, D.-W., Lukin, M.D., Demler, E.: Quantum Fluids of Self-Assembled Chains of Polar Molecules. Phys. Rev. Lett. {\bf 97}, 180413 (2006)


\bibitem{manybody7} Bruun G.M. and Taylor E.: Quantum Phases of a Two-Dimensional Dipolar Fermi Gas. Phys. Rev. Lett. {\bf 101}, 245301 (2008)
\bibitem{manybody9} Lutchyn R.M., Rossi E., Das Sarma S.: Spontaneous interlayer superfluidity in bilayer systems of cold polar molecules. Phys. Rev. A {\bf 82}, 061604(R) (2010)
\bibitem{manybody10} Cooper, N.R., Shlyapnikov, G.V.: Stable Topological Superfluid Phase of Ultracold Polar Fermionic Molecules. Phys. Rev. Lett. {\bf 103}, 155302 (2009)
\bibitem{manybody15} Carr S.T., Quintanilla J., Betouras J.J.: Lifshitz transitions and crystallization of fully polarized dipolar fermions in an anisotropic two-dimensional lattice. Phys. Rev. B {\bf 82}, 045110 (2010);
\bibitem{manybody16} Sun, K., Wu, C., Das Sarma, S.: Spontaneous inhomogeneous phases in ultracold dipolar Fermi gases. Phys. Rev. B {\bf 82}, 075105 (2010)
\bibitem{manybody17} Yamaguchi, Y., Sogo, T., Ito, T., Miyakawa, T.: Density-wave instability in a two-dimensional dipolar Fermi gas. Phys. Rev. A {\bf 82}, 013643 (2010)
\bibitem{manybody18} Pikovski, A.,Klawunn, M., Shlyapnikov, G.V., Santos, L.: Interlayer Superfluidity in Bilayer Systems of Fermionic Polar Molecules. Phys. Rev. Lett. {\bf 105}, 215302 (2010)
\bibitem{manybody22} Zinner, N.T., Wunsch, B., Pekker, D., Wang, D.W.: BCS-BEC Crossover in Bilayers of Cold Fermionic Polar Molecules. Phys. Rev. A. {\bf 85}, 013603 (2012)
\bibitem{manybody23} Zinner, N.T., Bruun, G.M.:Density Waves in Layered Systems with Fermionic Polar Molecules. Eur. Phys. J. D {\bf 65}, 133 (2011)
\bibitem{manybody24} Babadi, M., Demler, E.: Density ordering instabilities of quasi-two-dimensional fermionic polar molecules in single-layer and multi-layer configurations: exact treatment of exchange interactions. Phys. Rev. B {\bf 84}, 235124 (2011)
\bibitem{manybody25} Parish M.M. and Marchetti F.M.: Density instabilities in a two-dimensional dipolar Fermi gas. Phys. Rev. Lett. {\bf 108}, 145304 (2012)
\bibitem{manybody26} Sieberer, L.~M., Baranov, M.~A.: Collective modes, stability and superfluid transition of a quasi-two-dimensional dipolar Fermi gas. Phys. Rev. A {\bf 84}, 063633 (2011)
\bibitem{manybody27} Block J.K., Zinner N.T., Bruun G.M.: Density wave instabilities of tilted fermionic dipoles in a multilayer geometry. arXiv:1204.1822



\bibitem{dipcrystal1} Mora C., Parcollet O., Waintal X.: Quantum melting of a crystal of dipolar bosons. Phys. Rev. B {\bf 76}, 064511 (2007);
\bibitem{dipcrystal2} B{\"u}chler H.P. {\it et al.}: Strongly Correlated 2D Quantum Phases with Cold Polar Molecules: Controlling the Shape of the Interaction Potential. Phys. Rev. Lett. {\bf 98}, 060404 (2007)
\bibitem{dipcrystal3} Astrakharchik G.E., Boronat J., Kurbakov I.L., Lozovik Yu.E.: Quantum Phase Transition in a Two-Dimensional System of Dipoles. Phys. Rev. Lett. {\bf 98}, 060405 (2007)
\bibitem{wigner1934-2} Bonsall L. and Maradudin A.A.: Some static and dynamical properties of a two-dimensional Wigner crystal. Phys. Rev. B {\bf 15}, 1959 (1977)

\bibitem{santos2010} Klawunn, M., Duhme, J., Santos, L.: Bose-Fermi mixtures of self-assembled filaments of fermionic polar molecules. Phys. Rev. A {\bf 81}, 013604 (2010)
\bibitem{potter2010} Potter, A.~C. {\it et al.}: Superfluidity and dimerization in a multilayered system of fermionic polar molecules. Phys. Rev. Lett. {\bf 105}, 220406 (2010) 
\bibitem{wang2008a} Zhu K.-Y., Tan L., Gao X., Wang D.-W.: Quantum Fluids of Self-Assembled Chains of Polar Molecules at Finite Temperature. Chin. Phys. Lett. {\bf 25}, 48 (2008)

\bibitem{barbara2011} Capogrosso-Sansone B. and A. Kuklov A.: Superfluidity of flexible chains of polar molecules. J. Low Temp. Phys. {\bf 165}, 213 (2011)

\bibitem{ku2011} Ku M.J.H., Sommer A.T., Clark L.W., Zwierlein M.W.: Revealing the Superfluid Lambda Transition in the Universal Thermodynamics of a Unitary Fermi Gas. Science {\bf 335}, 563 (2012)

\bibitem{armstrong2011} Armstrong, J.R., Zinner, N.T., Fedorov, D.V., Jensen, A.S.: Analytic Harmonic Approach to the N-body problem. J. Phys. B: At. Mol. Opt. Phys. {\bf 44}, 055303 (2011)

\bibitem{ciach2011} Ciach A.: Simple lattice models of complex systems. Jour. Mol. Liquids {\bf 164}, 74 (2011)

\bibitem{artem2011d} Volosniev A.G., Armstrong J.R., Fedorov D.V., Jensen A.S., Zinner N.T.: Bound Chains of Tilted Dipoles in Layered Systems. Few-Body Syst. in press, arXiv:1112.2541

\bibitem{jeremy2011} Armstrong, J.R., Zinner, N.T., Fedorov, D.V., Jensen, A.S.: Quantum statistics and thermodynamics in the harmonic approximation. Phys. Rev. E {\bf 85}, 021117 (2012).

\bibitem{aymar2005} Aymar M. and Dulieu O.: Calculation of accurate permanent dipole moments of the lowest $^{1,3}\Sigma^+$ states of heteronuclear alkali dimers using extended basis sets. J. Chem. Phys. {\bf 122}, 204302 (2005).


\bibitem{kalia1981-1} Kalia R.K. and Vashishta P.: Interfacial colloidal crystals and melting transition. J. Phys. C {\bf 14}, L643 (1981)
\end{thebibliography}
\end{document}